\documentclass{aa}
\usepackage[dvips]{graphicx}
\usepackage{amsmath}
\def\simless{\mathbin{\lower 3pt\hbox
{$\rlap{\raise 5pt\hbox{$\char'074$}}\mathchar"7218$}}}   
\def\simmore{\mathbin{\lower 3pt\hbox
{$\rlap{\raise 5pt\hbox{$\char'076$}}\mathchar"7218$}}}   
\newcommand{\be}{\begin{equation}}

\newcommand{\bc}{\begin{center}}
\newcommand{\ec}{\end{center}}

\begin{document}
\title{Flares in GRB afterglows from delayed magnetic dissipation}
\titlerunning{Afterglow flares from delayed magnetic dissipation}
\author{Dimitrios Giannios}

\institute{Max Planck Institute for Astrophysics, Box 1317, D-85741 Garching, Germany}

\offprints{giannios@mpa-garching.mpg.de}
\date{Received / Accepted}

\abstract
{One of the most intriguing discoveries made by the {\it Swift} satellite is the flaring activity 
in about half of the afterglow lightcurves. Flares have been observed on both long and short duration GRBs and
on time scales that range from minutes to $\sim$1 day after the prompt emission. The rapid
evolution of some flares led to the suggestion that they are caused by late central engine 
activity. Here, I propose an alternative explanation that does not need reviving of the central engine. 
Flares can be powered by delayed magnetic dissipation in strongly magnetized (i.e. with initial 
Poynting to kinetic flux ratio $\simmore 1$) ejecta during its deceleration due to interaction with the external 
medium. A closer look at the length scales of the dissipation regions shows that magnetic dissipation can give 
rise to fast evolving and energetic flares. Multiple flares are also expected in the context of the model.

}

\maketitle

\section{Introduction} 
\label{intro}

A rich early afterglow phenomenology has recently been revealed thanks, to a large extent, to the X-ray
detector on board to the {\it Swift} satellite. The systematic study of X-ray afterglows (Zhang et al. 2006; 
Nousek et al. 2006) has shown that they are typically characterized by an initial steep decay, associated with 
the end of the prompt emission phase, followed by a flatter decay that lasts for about one hour and probably 
corresponds to the onset of the afterglow. At later times, 
the lightcurves are characterized by the ``normal'' decay observed before {\it Swift}.  

Another exciting {\it Swift} discovery is the afterglow flares observed in about 
half of the bursts (Burrows et al. 2005). Flares appear on different timescales $t_{\rm{f}}$ after the prompt emission 
ranging from  $\sim 100$ to $\sim 10^5$ sec, they are characterized by duration  $\delta t_{\rm {f}}$ such that
$\delta t_{\rm {f}}/t_{\rm{f}}\simless 1$, and show spectral evolution with respect to the smooth decay part of the 
afterglow. The flares are involving substantial energy release that ranges from a small fraction up to an amount equal 
to that of the prompt GRB emission (as in the case of  GRB050502b; Falcone et al. 2006) and they have been observed in the 
afterglow lightcurves of both long and short duration GRBs (Campana et al. 2006).

Although bumps in the afterglow lightcurves can be a result of slower shells catching
up with the decelerating ejecta (e.g. Rees \& M\'esz\'aros 1998) or interaction of the 
ejecta with a clumpy external medium, the rapid evolution of some of the flares has been used 
as an argument against their external shock origin. This led to the suggestion that the central engine revives
at later times (ranging from hundreds of seconds to days) giving rise to flares
through late internal shocks or magnetic dissipation (Burrows et al. 2005; Zhang et al. 2006). 
 Suggestions on how to revive the central engine have been made by King et al. (2005),
Dai et al. (2006), Fan et al. (2006), Perna et al. (2006).   

Here, I make an alternative suggestion which does not need late time activity 
of the central engine. I investigate the possibility that the deceleration of strongly 
magnetized ejecta leads to revival of MHD instabilities, that were delayed by time dilation before
the deceleration phase. The resulting dissipation produces flares in localized
reconnection regions in the flow.

\section{Magnetic dissipation in a decelerating shell}
\label{model}

Magnetic fields may be the main agent extracting energy from the central engine, leading to a Poynting-flux-dominated 
flow  (e.g., Thompson 1994; M\'esz\'aros \& Rees 1997; Spruit et al. 2001). Magnetic dissipation through 
instabilities in an axisymmetric flow (Lyutikov \& Blandford 2003; Giannios \& Spruit 2006) or directly through 
reconnection in a highly non-axisymmetric flow (Drenkhahn \& Spruit 2002) can power the prompt emission 
with high radiative efficiencies and accelerate the flow to ultrarelativistic speeds. 
The photospheric emission of a dissipative MHD flow is highly non-thermal and can be
responsible for the prompt GRB emission (Thompson 1994; Giannios 2006). 

Dissipation of Poynting flux is partial, however, if it depends on global magnetic instabilities. 
MHD instabilities (such as the kink instability) grow 
on the Alfv\'en crossing time in the fluid frame. Time dilation in the central engine frame 
by the bulk Lorentz factor of the flow $\Gamma$ results in slowing down of the dissipation in
ultrarelativistic flows. Giannios \& Spruit (2006) have shown that for large initial Poynting to 
kinetic flux ratios, the flow remains magnetically dominated
up to radii $\simmore 10^{16}$ cm, i.e. close to the distance where the deceleration of the flow 
against the external medium is expected.

The deceleration of the flow
can, however, naturally lead to reviving of the instabilities in the regions which come into causal 
contact again and to substantial magnetic dissipation in the afterglow phase. 
Another (and probably complementary)  possibility is that dissipation of magnetic energy is triggered 
by the crossing of the reverse shock through the magnetized shell (Thompson 2005).

Here, I investigate the possibility that late time dissipation takes place in this way and explore whether
it can power the commonly observed flares in the GRB afterglows. I discuss in particular the energetics and 
the variability expected from such events.     

\subsection{Deceleration of a magnetized shell}

The issue of flow acceleration and of prompt emission in a Poynting flux dominated flow
has been studied in a number of works in the context of GRB flows (e.g., Drenkhahn \& Spruit 2002; 
Vlahakis \& K\"onigl 2003; Giannios \& Spruit 2006).
At a sufficient distance from the source, the flow has accelerated to its terminal Lorentz
factor and the magnetic field is dominated by its toroidal component.
After the ultrarelativistic flow has swept up enough mass from the external medium 
(which may be the interstellar medium or the wind material), it starts decelerating,
driving a shock into the external medium. In this work, I focus on this deceleration phase.

I assume that a cold, strongly magnetized shell with a dominant toroidal magnetic field
is moving with initial bulk Lorentz factor $\Gamma_0\gg 1$ into the external medium. The magnetic 
content of the ejecta can be conveniently parameterized by the magnetization $\sigma$ defined as the 
ratio of Poynting to kinetic energy flux $\sigma =B^{\prime 2}/4\pi w$,
where $B'$ and $w=\rho c^2+e+p$ are the magnetic field and the enthalpy density  of the ejecta respectively
as measured by an observer comoving with the flow. The enthalpy density consists of the rest mass energy density $\rho c^2$, 
the energy density $e$ and the gas pressure $p$. Since the shell is assumed to be cold initially, 
its magnetization is $\sigma_0=B^{\prime 2}/4\pi \rho c^2$.    

The initial phase of the interaction of the magnetized shell with the
external medium has been studied by Zhang \& Kobayashi (2005) who have
shown that, for ``typical'' GRB parameters, a reverse shock forms
in the ejecta as long as $\sigma_0\simless 100$. In the shocked 
part of the ejecta, both gas and  magnetic field contribute
to the total pressure; their sum balances the pressure of the 
shocked external medium at the contact discontinuity. For high
enough initial magnetization of the ejecta $\sigma_0\simmore 0.1$, 
the magnetic pressure dominates the gas pressure in the shocked ejecta
irrespective of whether the reverse shock is Newtonian,  mildly relativistic or relativistic.
This can be shown by solving for the MHD shock conditions (Kennel \& Coroniti 1984; 
Zhang \& Kobayashi 2005), where it is also shown that shock compression leads to an increase of the 
magnetization of the ejecta, i.e. $\sigma_{\rm{sh}}\simmore \sigma_0$.

Thus, for $\sigma_0\simmore 0.1$, one can neglect the gas pressure 
of the shocked ejecta with respect to the magnetic pressure. The pressure balance at the contact discontinuity yields
\be
\frac{B^{\prime 2}_{\rm{sh}}}{8\pi}=\frac{4}{3}\Gamma_{\rm{sh}}^2\rho_{\rm{e}}c^2,
\label{bfield}
\end{equation}
where $B^{\prime 2}_{\rm{sh}}$, $\Gamma_{\rm{sh}}$ are magnetic field strength and the bulk Lorentz factor of the 
shocked ejecta  and $\rho_{\rm{e}}$ is the density of the 
external medium. From this point on, I focus on the shocked ejecta and omit the subscript
``sh'' for simplicity in the notation.   

\subsection{The magnetic dissipation region}

I have argued that the deceleration of the flow (and/or the crossing of the reverse shock)
revives MHD instabilities that lead to dissipation of magnetic energy through 
reconnection of magnetic field lines at different locations in the flow. The energy
released in a dissipative event and its observed duration depend on the characteristic
dimensions of the dissipation region. Let $l'_1$ and $l'_2$ be the characteristic  length scales in the 
radial and perpendicular to the bulk motion directions (measured by a comoving observer) of this
region respectively. Furthermore, I consider $l'_1$ to be the scale over which the field changes polarity
and that magnetic field lines are advected in the middle plane of this region where they reconnect. The 
reconnection plane is assumed to be  perpendicular to the radial direction.  
More generally, one should consider  an arbitrary angle between the radial direction and the normal to the
reconnection plane. However, I have checked that such a generalization only complicates the
analysis that follows, without introducing any essential changes on the results, and has been avoided.

The electromagnetic energy that is contained in the dissipation region is (in the central engine frame)
\be
E_{\rm{d}}=\Gamma l'_1l^{\prime 2}_2\frac{B^{\prime 2}}{4\pi}=\frac{8}{3}l'_1l^{\prime 2}_2\Gamma^3\rho_{\rm{e}}c^2,
\label{Epoynting}
\end{equation} 
where eq.~(\ref{bfield}) has been used in the last step. 
If a large fraction of the dissipated energy leads to fast moving electrons, they can efficiently radiate most of
it {\it if} the electrons' cooling timescale is shorter than the expansion time scale $t_{\rm{exp}}=r/\Gamma c$
of the flow. Here, I show that this is the case in the dissipation region under consideration.

An efficient cooling mechanism in the strongly magnetized flow is synchrotron cooling with a time
scale
\be
t_{\rm{syn}}=3m_{\rm{e}}c/4\sigma_{\rm{T}}\beta_{\rm{e}}^2\gamma_{\rm{e}}U_{\rm{B}},
\label{tsyn}
\end{equation}
where $\gamma_{\rm{e}}$ is the characteristic Lorentz factor at which electrons are accelerated in the
reconnection region and $U_{\rm{B}}=B^{\prime 2}/8\pi$ is the magnetic energy density. The available magnetic
energy per proton (and electron) is $\simmore m_p c^2$ for a flow with $\sigma_0\simmore 1$, which is the range
of $\sigma_0$ that I focus on in this study.
Dissipation of most of this energy to the electrons leads to $\gamma_{\rm{e}}\simmore m_{\rm{p}}/m_{\rm{e}}$.
By using, for an order of magnitude estimate, $\Gamma=50$, $r=3\cdot 10^{17}$ cm, $\rho_{\rm{e}}=10^{-24}$ gr cm$^{-3}$,
and $\gamma_{\rm{e}}=10^{3.5}$ in eqs.~({\ref{bfield}}) and (\ref{tsyn}), I find $t_{\rm{syn}}/t_{\rm{exp}}\sim 10^{-2}$
 and the characteristic synchrotron photon energy $E_{\rm{syn}}=3\Gamma \gamma_{\rm{e}}^2 q_{\rm e} \hbar B^{\prime}/ 
2m_{\rm e}c\sim 100$ eV, i.e. in the soft X-rays. Smaller values for  $t_{\rm{syn}}/t_{\rm{exp}}$ and higher for $E_{\rm{syn}}$ 
are found for ``typical'' parameters that correspond to the case where the external medium is the stellar wind. 

Thus, it is possible that a large fraction of the reconnected energy is promptly radiated away in the X-rays through
synchrotron emission.  This energy is emitted in the $\theta\sim 1/\Gamma$ forward cone, i.e., with a beaming factor 
$4\pi/ 2\pi (1-\cos \theta)\simeq 4\Gamma^2$. The beaming leads to isotropic equivalent energy $E_{\rm{f}}^{\rm{iso}}$ 
that is $4\Gamma^2$ times larger than the dissipated energy $E_{\rm d}$
\be
E_{\rm{f}}^{\rm{iso}}=4\Gamma^2E_{\rm d}.
\label{Eflare}
\end{equation}

If the released magnetic energy is to be observed as a flare, the dissipation must take place sufficiently fast. 
This leads to conditions on the length scales $l'_1$ and $l'_2$ that have to hold. These conditions are
investigated in the next section.

\subsection{Variability constraints}

In the dissipation region, magnetic fields are advected along a distance $\sim l'_1$ into
the reconnection plane which has a characteristic length $l'_2$. The observed duration
of the flare  $\delta t_{\rm{f}}$ is the sum of two contributions. The first is the ``radial'' duration
$\delta t_{\rm{r}}$ coming from the fact that while dissipation takes place, the flow  expands
radially resulting at different arrival times for photons that are emitted at different stages of the
dissipation event. The second contribution comes from the ``angular'' duration $\delta t_{\rm{ang}}$ that is related to the
delay in arrival of photons that are emitted in different locations on the reconnection plane.
The duration of the flare is $\delta t_{\rm{f}}= \delta t_{\rm{r}}+\delta t_{\rm{ang}}$. I turn to these
two contributions and study them separately.

The ``radial'' duration  of the flare depends on how fast reconnection takes place. 
The speed $v_{\rm{r}}$ at which magnetic reconnection proceeds in a strongly magnetized plasma is enhanced
by the relativistic kinematics of the problem (e.g., Lyutikov \& Uzdensky 2003; Lyubarsky 2005)
and is a substantial fraction $\varepsilon \sim 0.1$ of the Alfv\'en speed which is close to the speed of light 
for $\sigma_0\simmore 1$. Magnetic field lines are advected for a distance $\l'_1/2$ above and below the reconnection plane 
with speed $v_{\rm{r}}$, which leads to a comoving dissipation timescale $t'_{\rm{d}}\simeq l'_1/(2v_{\rm{r}})\simeq l'_1/
(2\varepsilon c)$. Dividing this timescale with the expansion timescale $r/\Gamma c$, one has an estimate on the 
fractional duration $\delta t_{\rm{r}}/t_{\rm{f}}$ of the flare because of the radial motion of the reconnecting 
region
\be
\delta t_{\rm{r}}/t_{\rm{f}}=\Gamma l'_1/2\varepsilon r.
\label{tr}
\end{equation}
   
The ``angular'' duration is related to the characteristic length $l'_2$ of the reconnecting region.
A first constraint to $l'_2$ comes from causality arguments that yield $l'_2\simless r/\Gamma$. The observer sees an emitting region
with length $l'_2$ from a characteristic angle $\sim 1/\Gamma$. The difference in the arrival time of two photons 
coming from the two edges of the emitting region is $\delta t_{\rm{ang}}\simeq l'_2/c\Gamma$.
How the observer time is related to the radius and bulk Lorentz factor of the blast wave depends on the details of 
the deceleration profile which, in turn, depend on the density profile of the external medium and on the magnetic content
of the ejecta. To keep the study general, I assume that $t_{\rm{f}}= r/\alpha c\Gamma^2$, where
$\alpha$ is a parameter with typical values $\sim$ a few  [e.g., for baryonic ejecta decelerated in constant
density medium $\alpha=4$ (Waxman 1997) and in stellar wind $\alpha=2$ (Pe'er \& Wijers 2005)]. 
Using the expressions for $\delta t_{\rm{ang}}$ and  $t_{\rm{f}}$, I have
\be
\delta t_{\rm{ang}}/t_{\rm{f}}=\alpha \Gamma l'_2/r.
\label{tang}
\end{equation}    

The observed duration of the flare $\delta t_{\rm{f}}$ is given by the sum of the radial and
angular timescales. Using eqs. (\ref{tr}) and (\ref{tang}), I arrive at 
\be
\delta t_{\rm{f}}/t_{\rm{f}}=\Gamma l'_1/2\varepsilon r+\alpha \Gamma l'_2/r.
\label{tflare}
\end{equation}
The observed fractional duration of the flare thus constrains both 
characteristic length scales of the reconnection region
\begin{align}
\label{con1} 
l'_1&\simless 2\varepsilon r (\delta t_{\rm{f}}/t_{\rm{f}})/\Gamma,
\\[1mm]
\label{con2}
l'_2&\simless r (\delta t_{\rm{f}}/t_{\rm{f}})/\alpha \Gamma.
\end{align}

In the process of deriving the angular time scale, I have implicitly ignored fast fluid motions
in the frame moving with the bulk of the flow. The material leaves the
reconnection region with the Alv\'en speed (e.g. Lyubarsky 2005) however, which is close to the speed
of light. This can affect the estimate (\ref{con2}) by allowing for larger values of
$l'_2$ (and ultimately for larger energy release in a flare of a specific 
$\delta t_{\rm{f}}/t_{\rm{f}}$; see the rest of the section). 
This effect is particularly pronounced in $\sigma_0 \gg 1$ flows which can contain
emitting regions with high ``internal'' Lorentz factors and lead to more rapid variability
than that estimated here (Lyutikov \& Blandford 2003; Lyutikov 2006). 

Expressions (\ref{con1}) and (\ref{con2}), combined with eqs. (\ref{Eflare}) and (\ref{Epoynting}),
yield a limit on the isotropic equivalent energy for a flare that can be produced by a single reconnection event
\be
E_{\rm{f}}^{\rm{iso}}\simless 20\varepsilon (\delta t_{\rm{f}}/t_{\rm{f}})^3\Gamma^2r^3 \rho_{\rm {e}}c^2/\alpha^2,
\label{Ep}
\end{equation}
where the numerical factor of the last expression 64/3 was set to 20.
Eq.~(\ref{Ep}) can be rewritten in a more compact form by using the fact that 
$4\pi r^3\rho_{\rm{e}}/3\simeq M_{\rm{swept}}$, where $M_{\rm{swept}}$ is the mass of the external medium that
has been swept up by the blast wave. Furthermore, the forward shock conditions dictate that $\Gamma^2 M_{\rm{swept}}c^2$
is to be identified with the energy that has been passed onto the forward shock $E_{\rm{fs}}^{\rm{iso}}$.
Expression (\ref{Ep}) can, thus, be rewritten 
\be
E_{\rm{f}}^{\rm{iso}}\simless  5\varepsilon (\delta t_{\rm{f}}/t_{\rm{f}})^3E_{\rm{fs}}^{\rm{iso}}/\alpha^2.
\label{Ef}
\end{equation}
 
From the last expression, it is clear that the model predicts that the energy available to power the
fastest evolving flares with $\delta t_{\rm{f}}/t_{\rm{f}}\sim 0.1$ is 3 orders of magnitude less than the energy
that can be released during a smoother flare with $\delta t_{\rm{f}}/t_{\rm{f}}\sim 1$. The energy of the
flare depends on  $E_{\rm{fs}}^{\rm{iso}}$, the radial dependence of which has not been studied for
strongly magnetized ejecta. The key open question is how fast is the Poynting flux passed
onto the forward shock. This has an important effect on the bulk Lorentz factor of the flow as a function of radius 
$\Gamma (r)$ (see Zhang \& Kobayashi 2005; Lyutikov 2006) and, therefore, on the (observer) timescales over which 
magnetic dissipation can result in powerful flares. A detailed study of the dynamics of the deceleration of ejecta with 
$\sigma_0\simmore 1$ can settle these issues and yield valuable constraints on the  model. 

Nevertheless, one can have a rough estimate of the energetics of a flare by assuming that the energy in the forward shock 
at some radius $r$ is a large fraction of the energy that was initially carried by the ejecta $E_{\rm{ej}}^{\rm{iso}}$. 
In view of eq.~(\ref{Ef}), it means that a fraction
\be
f\simless 5\varepsilon (\delta t_{\rm{f}}/t_{\rm{f}})^3/\alpha^2
\label{frac}
\end{equation} 
of the initial blast wave energy can power a flare of duration $\delta t_{\rm{f}}$  at (observer) time
$t_{\rm{f}}$. As an arithmetic example, I set $\alpha^2\simeq 10$ and $\varepsilon \simeq 0.2$, from which I find  
$f\simless 0.1(\delta t_{\rm{f}}/t_{\rm{f}})^3$. One can go a step further and compare  the 
energy of the flare with that of the prompt GRB emission by assuming that the 
latter is a fraction $\eta \sim 0.1$ of the energy of the blast wave. This means that a flare coming from 
late magnetic dissipation can reach a fraction $\sim  (\delta t_{\rm{f}}/t_{\rm{f}})^3$ of the prompt 
GRB emission. So the model may explain powerful and fast evolving 
flares at the same time. On the other hand, there is a clear prediction that fast evolving flares are less
energetic than the smoother ones.      

\subsection{Multiple flares}

Very often, the afterglow light curves are characterized by multiple flares.
In the context of this model, this corresponds to multiple reconnection regions. 
The physical scales of such region are constrained by eqs.~(\ref{con1}) and (\ref{con2}) and
one can check that they correspond to a moderate fraction of the volume of a shell
with $1/\Gamma$ opening angle that emits toward the observer. So, a single flare 
does not use up all the Poynting flux in the observer's light cone and repetition
is possible. It is even likely that dissipation in one region can induce instabilities
and magnetic dissipation in neighboring regions, which results in a sequence of flares.

\section{Conclusions}

The afterglow flares observed by {\it Swift} have been suggested to indicate late time activity 
of the central engine. As an alternative, I  propose here that flares are produced in the deceleration phase
of strongly magnetized ejecta. This comes out naturally from Poynting models where 
instabilities give partial magnetic dissipation (prompt emission and acceleration; see, e.g., Giannios \& Spruit 2006). 
For high bulk Lorentz factor of the flow, the instabilities slow down because of the time dilation effect
and the flow remains strongly magnetized in the afterglow phase. The deceleration of the flow
due to interaction with the external medium (and possibly the crossing of the reverse shock; 
Thompson 2005) revives the instabilities and leads to delayed magnetic dissipation. 

I have looked at the magnetic energy available in the flow to power such flares and how
it depends on the length scales of the dissipation region. For a flare to have
the observed rapid evolution, the scales of this region cannot be arbitrarily large, and this constrains 
the available energy per flare. I estimate
that flares caused by delayed reconnection can be powerful for both long and short GRBs, with isotropic equivalent 
energies up to a fraction $\sim 0.1 \cdot (\delta t_{\rm{f}}/t_{\rm{f}})^3$ of the blast wave energy. Although this
estimate contains uncertainties related to the speed of magnetic reconnection and the deceleration profile 
of the ejecta [see eq.~(\ref{frac})], it indicates that powerful flares through delayed 
magnetic dissipation are possible and that smooth flares are expected to be more energetic than spiky ones.
Multiple flares are expected in this model as a result of dissipation in multiple neighboring 
regions in the decelerating flow. Furthermore, inverse Compton scattering of the flare photons that takes 
place in the forward shock can result in GeV-TeV flares as shown by Wang et al. (2006).

\begin{acknowledgements}
I thank Henk Spruit and Nick Kylafis for valuable suggestions and discussions.
I acknowledge support from the EU FP5 Research Training Network ``Gamma Ray Bursts:
An Enigma and a Tool.'' 
\end{acknowledgements}


\begin{thebibliography}{}

\bibitem{} Burrows, D.~N., Romano, P., Falcone, A.~D., et al. 2005, Science, Vol. 309, Issue 5942, 1833 
\bibitem{} Campana, S., Tagliaferri, G., Lazzati, D., et al. 2006, A\&A, in press (astro-ph/0603475)
\bibitem{} Dai, Z.~G., Wang, X.~Y., Wu, X.~F., \& Zhang, B. 2006, Science, 311, 1127  
\bibitem{} Drenkhahn, G., \& Spruit, H.~C. 2002, A\&A, 391, 1141
\bibitem{} Falcone, A.~D., Burrows, D.~N., Lazzati, D.,  et al. 2006, ApJ, 641, 1010
\bibitem{} Fan, Y. Z., Zhang, B., \& Proga, D. 2006, ApJ, 635, L129 
\bibitem{} Giannios, D., \& Spruit, H.~C. 2006, A\&A, 450, 887
\bibitem{} Giannios, D. 2006, A\&A, submitted (astro-ph/0602397)
\bibitem{} Kennel, C.~F., \& Coroniti, F.~V. 1984, ApJ, 283, 694
\bibitem{} King, A., O'Brien, P.~T., Goad, M.~R., Osborne, J., Olsson, E., \& Page, K. 2005, 630, L113
\bibitem{} Lyubarsky, Y.~E. 2005, MNRAS, 358, 113
\bibitem{} Lyutikov, M., \& Uzdensky, D. 2003, ApJ, 589, 893
\bibitem{} Lyutikov, M., \& Blandford R.~D. 2003, astro-ph/0312347
\bibitem{} Lyutikov, M. 2006, New Journ. of Phys., in press (astro-ph/0512342) 
\bibitem{} M\'esz\'aros, P., \& Rees, M.~J. 1997, ApJ, 482, L29
\bibitem{} Pe'er, A., \& Wijers, R.~A.~M.~J. 2005, ApJ, submitted (astro-ph/0511508)
\bibitem{} Perna, R., Armitage, P.~J., \& Zhang, B. 2006, ApJ, 636, L29
\bibitem{} Rees, M.~J., \& M\'esz\'aros, P. 1998, ApJ, 496, L1
\bibitem{} Spruit, H.~C., Daigne, F., \& Drenkhahn, G. 2001, A\&A, 369, 694
\bibitem{} Nousek, J.~A., Kouveliotou, C., Grupe, D., et al. 2006, ApJ, 642, 389 
\bibitem{} Thompson, C. 1994, MNRAS, 270, 480
\bibitem{} Thompson, C. 2005, ApJ, in press (astro-ph/0507387)
\bibitem{} Vlahakis, N., \& K\"onigl, A. 2003, ApJ, 596, 1080
\bibitem{} Wang, X.~Y., Li, Z., \& M\'esz\'aros, P. 2006, ApJ, 641, L89 
\bibitem{} Waxman, E. 1997, ApJ, 491, L19
\bibitem{} Zhang, B., \& Kobayashi, S. 2005, ApJ, 628, 315 
\bibitem{} Zhang, B., Fan, Y.~Z., Dyks, J., et al. 2006, ApJ, 642, 354 


\end{thebibliography}
\end{document}